# Protogalactic evolution and magnetic fields

H. Lesch[1] and M. Chiba[1,2]

[1] Max-Planck-Institut für Radioastronomie, Auf dem Hügel 69, D-53121 Bonn, Germany
[2] Astronomical Institute, Tohoku University, Aoba-ku, Sendai 980-77, Japan

**Abstract.** We show that the relatively strong magnetic fields ($\geq 1\mu$G) in high redshift objects can be explained by the combined action of an evolving protogalactic fluctuation and electrodynamic processes providing the magnetic seed fields. Three different seed field mechanisms are reviewed and incorporated into a spherical "top-hat" model and tidal torque theory for the fate of a forming galaxy in an expanding universe. Very weak fields $10^{-19} \sim 10^{-23}$G created in an expanding overdense region are strongly enhanced due to the dissipative disk formation by a factor $\sim 10^4$, and subsequently amplified by strong non-axisymmetric flow by a factor $\sim 10^{6-10}$, depending on the cosmological parameters and the epoch of galaxy formation. The resulting field strength at $z \sim 0.395$ can be of the order of a few $\mu$G and be close to this value at $z \sim 2$.

## 1. Introduction

The origin of galactic magnetic fields remains enigmatic (Rees 1987; Kronberg 1994). Several models have been proposed, such as dynamo amplification of a seed magnetic field (e.g. Zeldovich, Ruzmaikin & Sokoloff 1983). Dynamo theories, however have difficulties to explain the presence of strong magnetic fields in high redshift objects, since at least a couple of billion years are required for the very weak seed fields ($\sim 10^{-20}$G) to be sufficiently amplified. For example the damped Ly$\alpha$ systems at a redshift $z \sim 2$ have undergone about 1 Gyr evolution, yet posses magnetic fields of order of a few $\mu$G (Wolfe et al. 1992, see other examples by Kronberg & Perry 1982 and Kronberg et al. 1992 at $z \sim 0.395$). Although these results have been criticized from the viewpoint of interpreting observational data (Perry, Watson & Kronberg 1993), the very possibility of the presence of $\mu$G magnetic fields at these very early stages of galactic evolution deserves special attention, since such field strengths cannot be explained by standard dynamo models. In addition to observational difficulties Kulsrud & Anderson (1992) argue on theoretical grounds that strengths and coherence lengths of large-scale magnetic fields cannot be produced via dynamo amplification.

Chiba & Lesch (1994) developed a model for magnetic field amplification in disk galaxies by taking into account the action

*Send offprint requests to*: H. Lesch

of non-axisymmetric disturbances on the velocity field of the gas. They found that exponential growth of magnetic fields is easily achieved on very short time scales of the order of the rotation time if spiral arms or bars are excited in the disk. It is the aim of this contribution to show that an extended version of this model explains the presence of strong magnetic fields at high redshift.

Since any magnetic field amplification needs a seed field whose strength is enhanced by magnetohydrodynamical processes, we start in Sect.2 with the description of three mechanisms which can provide the seed fields either by taking into account the different reaction of electrons and protons to an intense background radiation field (Compton induced fields), by considering the difference in mobility of electrons and protons in a gravitational field of a rotating gas (Biermann's battery) or by investigating the friction between neutral and ionized gas in a galactic shear flow. In Sect.3. we present a general model for the evolution of a protogalactic fluctuation on its way towards a disk galaxy in terms of a spherical "top-hat" model and tidal torque theory (Peebles 1980). The final amplification of the seed fields by a dissipative galaxy collapse and by evolving nonaxisymmetric structures (spirals and bars) in a forming galactic disk is described in Sect.4. Our conclusions are given in Sect.5.

## 2. Origin of magnetic fields in protogalaxies

The amplification of magnetic fields in conducting media is described by the hydromagnetic equation,

$$\frac{\partial \mathbf{B}}{\partial t} = \nabla \times (\mathbf{V} \times \mathbf{B}) - \nabla \times \eta(\nabla \times \mathbf{B}). \tag{1}$$

$\eta$ is the diffusivity of the plasma and $\mathbf{V}$ the plasma velocity. It is the central problem of the magnetohydrodynamic theory to find the appropriate galactic velocity fields which can amplify magnetic fields in the presence of diffusion. There is no source term in eq.(1). That is to say, there is no outright creation of field in the hydromagnetic description of galactic magnetic fields. Hence, if at any point in time the universe was devoid of magnetic fields, then as far as hydromagnetic effects are concerned, there would be no magnetic field at any other time. Biermann (1950) made the point, that there are thermal and inertial "battery" effects in a moving ionized gas to create magnetic fields. The essence of any battery process is that currents are produced whenever the mean velocities of negative charge carriers and positive charge carriers differ. In general, the negative charge carriers are electrons and as such they are orders of magnitudes less massive than the positive charge carriers. This makes electrons much more responsive to inertial drag forces than the ions are. The combination of a gravitational field with differential rotation leads to a nonconservative force acting essentially upon the electrons. These two ingredients occur naturally in disk systems with a central radiation source, as well as in stars, as was first pointed out by Biermann (1950). The ions are concentrated to the equatorial plane by the generated electric field. However, this field cannot cancel the centrifugal acceleration completely, i.e the charges must move and meridional currents appear. The resulting magnetic field of a spherical over-dense region in an expanding universe, with an expansion rate $a$ of universe is governed by the following equation (e.g. Biermann 1950; Zeldovich et al. 1983)

$$\frac{1}{a^2}\frac{\partial}{\partial t}a^2 B = \frac{m_i c}{2e}\nabla \times \mathbf{g} \simeq \frac{m_i c}{2e}\omega^2. \quad (2)$$

$g$ denotes the centrifugal acceleration and $\omega$ is the rotation velocity. $m_i$ is the ion mass, c the velocity of light and e is the charge (all in cgs units).

The expansion rate is given by $a = 1/(1 + z)$. The tidal torque theory for the origin of galaxy angular momentum (Peebles 1969) shows that in the linear regime the density perturbations grow $\propto t^{2/3}$ and the angular velocity changes as $\omega \sim t^{1/3} \sim (1+z)^{-1/2}$. Furthermore, after the perturbation is well separated from the background universe and collapsed, a galaxy gains torque proportional to $Q/t^2$, where $Q$ is the quadrupole moment $\sim MR^2$. Most of angular momentum is transferred when a galaxy turns around at the epoch $z_{turn}$ (when the density reaches about 5.6 times the value of the background universe) with a radius $R_m$. Since after $z_{turn}$ a sphere collapses and virializes into a radius $R_v$, which is about a half of the maximum radius $R_m$ (see next section for details), and since a sphere is spun-up by a factor $(R_m/R_v)^2 \sim 4$, more than that due to the torque gained after the system is bounded, we postulate that the angular velocity $\omega$ before $z_{turn}$ follows the equation, $\omega = \frac{\omega_v}{4}\left(\frac{1+z}{1+z_{turn}}\right)^{-1/2}$, where $\omega_v$ is the angular velocity at the epoch of virialization, say $z_{vir}$.

Thus we can obtain the dependence of the magnetic field strength on the redshift (before $z_{turn}$)

$$B(z) = \frac{m_i c \omega_v^2}{32 e H_0}(1+z)^2(1+z_{turn})\int_z^{z_{rec}}\frac{dz}{(1+z)^5(1+\Omega z)^{1/2}}, (3)$$

where $\Omega$ is the cosmological density parameter. $z_{rec}$ denotes the redshift of recombination (about $10^3$). With the Hubble constant in units of $h_{50}$ and $z \sim z_{turn}$, we arrive at a magnetic field at the turnover time

$$B(z_{turn}) \sim 2.5 \times 10^{-23} h_{50}^{-1}\left(\frac{\omega_v}{2\times 10^{-17}\,s^{-1}}\right)^2\left(\frac{\Omega}{0.1}\right)^{-1/2}$$
$$\times \left(\frac{1+z_{turn}}{8}\right)^{-3/2} G. \quad (4)$$

This form suggests that the later turn-over (smaller $z_{turn}$) sphere gives stronger magnetic fields as noted by Pudritz & Silk (1989).

A second battery process was invoked by Mishustin & Ruzmaikin (1973). They considered the interaction of a rotating electron-proton plasma with the intense background radiation. Thermal electrons scatter the photons of the background radiation via Compton scattering and gain energy and momentum, thereby drifting relative to the protons, i.e. producing a current. This current induces a magnetic field, whose time evolution is described by

$$\frac{1}{a^2}\frac{\partial}{\partial t}a^2 B = \frac{m_e c}{e}\frac{2\omega}{\tau_\gamma}, \quad (5)$$

where $m_e$ is the electron mass and $\tau_\gamma$ denotes the optical depth for Compton scattering, which is a sensitive function of the redshift, $\tau_\gamma = 3m_e c/4\sigma_T \rho_\gamma(0)(1+z)^4 \equiv \tau_\gamma(0)(1+z)^{-4}$. $\sigma_T$ is the Thomson cross section and $\rho_\gamma(0) \simeq 4\cdot 10^{-13}$ ergs cm$^{-3}$ is the present energy density of the background radiation. The magnetic field at $z \simeq z_{turn}$ is then

$$B(z_{turn}) \sim 2.2 \times 10^{-23} h_{50}^{-1}\left(\frac{\omega_v}{2\times 10^{-17}\,s^{-1}}\right)\left(\frac{\Omega}{0.1}\right)^{-1/2}$$
$$\times \left(\frac{1+z_{turn}}{8}\right)^{5/2} G. \quad (6)$$

Last, but not least, we describe the self-generation of magnetic fields by sheared flows in weakly ionized plasmas proposed by Huba & Fedder (1993). Again, the different mobility of electrons and ions is used. The electrons collide with neutral atoms, thereby drifting relative to the ions. If the global system rotates differentially, the drift corresponds to a current, which induces a magnetic field. The field generation term is

$$\frac{1}{a^2}\frac{\partial}{\partial t}a^2 B = \frac{m_e c}{e}\nabla \times \nu_{en}(\mathbf{V_i} - \mathbf{V_n}) \simeq \frac{m_e c}{e}\nu_{en}\frac{v_r}{l_{shear}}. \quad (7)$$

$\nu_{en}$ denotes the electron-neutral collision frequency, $v_r$ is the relative ion-neutral drift speed, and $l_{shear}$ is the shear length. $\mathbf{V_i}$ is the ion fluid velocity and $\mathbf{V_n}$ is the neutral fluid velocity. Supposing that $l_{shear}$ and $v_r$ are constant for simplicity, the magnetic field at $z \simeq z_{turn}$ is then

$$B(z_{turn}) \sim 4.6 \times 10^{-19} h_{50}^{-1}\left(\frac{\nu_{en}}{10^{-9}\,s^{-1}}\right)\left(\frac{\Omega}{0.1}\right)^{-1/2}$$
$$\times \left(\frac{1+z_{turn}}{8}\right)^{-3/2}\left(\frac{v_r}{0.01 km s^{-1}}\right)\left(\frac{l_{shear}}{1 kpc}\right)^{-1} G. (8)$$

The electron-neutral collision frequency was estimated for a gas temperature of $10^4$ K and a density of about 1cm$^{-3}$ (Spitzer 1968).

It turns out that the magnetic fields created during the expansion of an over-dense sphere are very weak, $10^{-19} \sim 10^{-23}$ G, compared to the values at the present and $z \sim 2$ (Rees 1987). We need further mechanisms to increase such weak fields drastically as will be described below.

## 3. Disk formation and magnetic fields

After the turn-around epoch $z_{turn}$, the sphere breaks away from the general expansion, and starts to collapse. The crossing among each spherical shell and the development of large fluctuations inside the sphere will result in the violent relaxation, leading to the virial equilibrium at the epoch $z_{vir}$. According to the virial theorem, if the dissipative force is small (see below for baryons), a half of the potential energy at $z_{turn}$ must be converted to the kinetic energy at $z_{vir}$, so that the radius in the equilibrium, $R_v$, is about a half of the maximum radius $R_m$ at $z_{turn}$ (Peebles 1980). Thus, the seed magnetic fields created at earlier epochs are strengthened by a factor $(R_m/R_v)^2 \sim 4$, because the magnetic fluxes are well conserved due to the high conductivity.

The virialized sphere contains both collisionless dark-matter particles and roughly 10% baryonic matter. The dissipation is inevitable in baryons, and thus the baryonic matter sinks dissipatively by about a factor $F^{-1} \sim 10$, where $F$ is the baryonic fraction in the sphere (Blumenthal et al. 1986; Barnes 1987). The radius of the baryonic sphere, $R_D$, is therefore $\sim FR_v$, so that the strength of magnetic fields is further enhanced by about a factor $F^{-2} \sim 100$.

The tidal torque theory in galaxies suggests that the sphere gains angular momentum by tidal torque, represented by a spin parameter $\lambda \equiv J|E|^{1/2}G^{-1}M^{-5/2} \sim 0.07$, where $J$ is the angular momentum, $E$ the total energy, and $M$ the total mass (Peebles 1969). As the collapse proceeds, the sphere is spun-up because the angular momentum is roughly conserved (depending on the degree of substructuring, Katz & Gunn 1991). Thus, the collapse becomes anisotropic, leaving a rotationally supported and approximately exponential disk in surface density (Fall & Efstathiou 1980). The vertical scale height of a protogalactic disk, $H$, is roughly given as $H \sim c_s/\omega$, where $c_s$ is the sound velocity. The rotational frequency $\omega$ is essentially determined by dark matter, which does not drastically change in the disk-formation process except for the inner region where the self-gravity of a disk is important (*cf.* Saio & Yoshii 1990). However, the sound velocity, determined by the interplay between radiative cooling and supernova heating, would change with time in systematic ways, which are beyond the scope of this article. In order to get the approximate value for $H$, we utilize the scale height of old disk stars in the Galaxy, because these stars would have formed in the early stage. We adopt $H \sim 500pc$, which is intermediate between scale heights of a thin disk ($\sim 350pc$) and of a thick disk ($\sim 1000pc$) (Freeman 1987). Turning to the magnetic fields, radio observations suggest that strong fields are mainly parallel to a disk plane except for a galactic center (Sofue et al. 1986). Thus, the magnetic fields are compressed and strengthened along a disk by at least a factor $(R_D/H) \sim (20kpc/500pc) \sim 40$.

Therefore, after the sphere turns around at $z_{turn}$ with radius $R_m$, it collapses into a disk with a half thickness $H$ and radius $R_D$, at the epoch of disk formation, say $z_f$, so that the magnetic-field strength is strongly increased by roughly a factor $f_c$, where

$$f_c \sim 1.6 \times 10^4 \left(\frac{R_m/R_v}{2}\right)^2 \left(\frac{0.1}{F}\right)^2 \left(\frac{R_D/20kpc}{H/500pc}\right). \quad (9)$$

The magnetic-field strength at $z_f$ is thus given as $B(z_f) \sim f_c B(z_{turn})$, so that,

$$B(z_f) = \begin{cases} 3.6 \times 10^{-19} h_{50}^{-1} \left(\frac{\omega_v}{2 \times 10^{-17} s^{-1}}\right) \left(\frac{\Omega}{0.1}\right)^{-1/2} \\ \quad \times \left(\frac{1+z_{turn}}{8}\right)^{5/2} (G) : \text{ for Compton,} \\ 4.0 \times 10^{-19} h_{50}^{-1} \left(\frac{\omega_v}{2 \times 10^{-17} s^{-1}}\right)^2 \left(\frac{\Omega}{0.1}\right)^{-1/2} \\ \quad \times \left(\frac{1+z_{turn}}{8}\right)^{-3/2} (G) : \text{ for Battery,} \\ 7.3 \times 10^{-15} h_{50}^{-1} \left(\frac{\nu_{en}}{10^{-9} s^{-1}}\right) \left(\frac{v_r}{0.01 km s^{-1}}\right) \\ \quad \times \left(\frac{l_{shear}}{1kpc}\right)^{-1} \left(\frac{\Omega}{0.1}\right)^{-1/2} \left(\frac{1+z_{turn}}{8}\right)^{-3/2} (G) : \\ \text{for Shear.} \end{cases} \quad (10)$$

We emphasize that *the usually adopted strength of seed magnetic fields is greatly strengthened by a factor $\sim 10^4$ due to the galaxy collapse*. This point has been always missed in the published papers concerning the estimation of magnetic fields in protogalaxies (e.g. Zeldovich et al. 1983). Note here that although we have considered the galaxy collapse step by step, the actual processes of virialization, dissipation, and disk formation may occur simultaneously (Katz & Gunn 1991). However, the total collapse factor $\sqrt{f_c}$ of a disk is not greatly affected by such considerations. We note that the collapse into a disk is not uniform, so that there are strong density perturbations in a collapsed disk (Katz & Gunn 1992): the magnetic field would be locally enhanced much more by orders of magnitude. Also, plasma processes described before to create fresh magnetic fields may still work in the collapse stage. However, these contributions do not increase the field strength by more than a factor three; the collapse is the overwhelming factor.

## 4. Field amplification and dynamics

A protogalactic disk settles into a centrifugal equilibrium and star formation proceeds. According to the numerical simulation by Katz & Gunn (1991), strong non-axisymmetric structures such as bars and spirals have emerged in a collapsing disk. The gravitational torques among substructures lead to angular-momentum exchange and thus trigger the mass re-distribution via radial gas flows.

Evolution of magnetic fields in strongly non-axisymmetric galaxies has been investigated by Chiba & Lesch (1994). The induced radial flow due to angular-momentum transport and velocity shear of galactic rotation result in the rapid shearing of the radial component $B_r$ into the azimuthal component $B_\phi$, producing magnetic fields with time scale $(\partial v_r/\partial r)^{-1} \sim 2\pi/\omega$ in a gas-rich disk, where $v_r$ is the radial velocity. The mechanism involved is invoking the coupling between radial flow and differential rotation, not the single gas (or magnetic-field) compression. Furthermore, it is found that highly non-axisymmetric component of gas flow leads to the exponential and oscillatory growth of magnetic fields by driving a resonant magnetic wave with a background spiral or bar density wave, and the time scale is of the order of the inverse of the spiral- or bar-forcing frequency $1/2(\omega - \omega_p)$ ($\omega_p$ is the angular pattern speed of of a bar or spiral disturbance). We refer Fig.5 in Chiba & Lesch (1994) to illustrate how this resonance is working in a bar system. Thus the characteristic time of the field growth $\tau$ is well given as the dynamical time scale, i.e.

$$\tau = min(\tau_{circ}, 1/2(\omega - \omega_p)), \quad (11)$$

where $\tau_{circ}$ is the orbital time of circular velocity. If we consider the vertical gas flow by young stars or supernovae, or spatially varied dissipation of magnetic fields, the field components along a disk can be coupled to the vertical component, so that three dimensional fields will not result in ultimate decay (see Chiba & Lesch 1994 for details). Magnetic seed fields which are created and enhanced in a protogalactic disk are amplified exponentially with time scale $\tau$. The field strength at redshift $z$ is then given by,

$$B(z) \sim B(z_f) \exp\left[\frac{19.6 h_{50}^{-1}(Gyr)}{\tau(Gyr)} \int_z^{z_f} \frac{dz}{(1+z)^2(1+\Omega z)^{1/2}}\right]$$
$$\equiv B(z_f) f_a(z; z_f, \Omega), \quad (12)$$

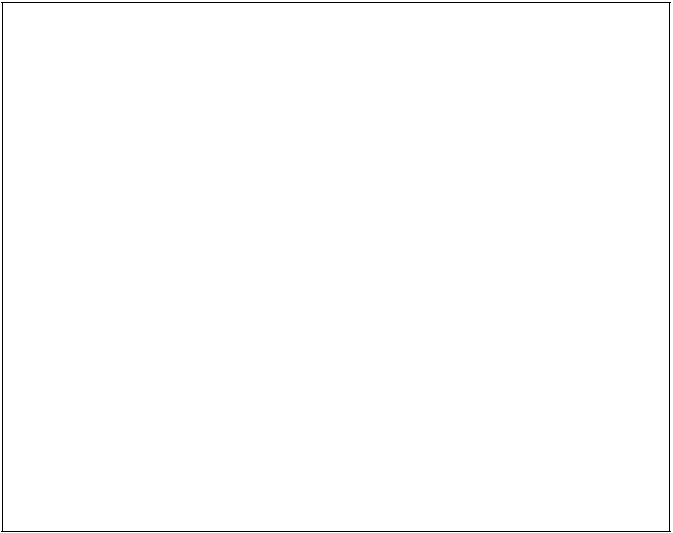

**Fig. 1.** The amplification factor, $f_a$, given in Eq.(12), as a function of the redshift, $z$, when $\tau = 0.1$ Gyr. $z_f$ denotes the epoch of disk formation. Solid lines are for $\Omega = 0.1$, while dotted ones for $\Omega = 1.0$.

Figure 1 shows the amplification factor $f_a$ as a function of $z$ ($\tau = 10^8$ yr). Supposing the flat universe, $\Omega = 1$, and $z_f \sim 3$ as a disk formation epoch, the amplification factor at $z \sim 2$ for observed damped Ly$\alpha$ clouds is then $f_a \sim 10^4$, so that $B(z \sim 2) \sim 10^{-15}$G (Compton), $10^{-15}$G (Battery), and $10^{-11}$G (Shear). If $z_f \sim 4$, we get $f_a \sim 10^6$, so that $B(z \sim 2) \sim 10^{-13}$G (Compton), $10^{-13}$G (Battery), and $10^{-9}$G (Shear). If $z \sim 0.395$ (Kronberg et al. 1992), $B(z \sim 0.395)$ is well larger than $\mu$G. If we adopt $\Omega = 0.1$, the value supported by recent studies of observational cosmology (e.g. Fukugita et al. 1990), the look-back time from the epoch of an observed galaxy $z$ to the formation epoch $z_f$ is longer so that the value of $f_a$ becomes larger. Actually for $(z, z_f) = (2, 3)$, we get $f_a \sim 10^6$ (suggesting $B = 10^{-9} \sim 10^{-13}$), and for $(z, z_f) = (2, 4)$, $f_a \sim 10^{10}$ (suggesting $B = 10^{-5} \sim 10^{-9}$). It suggests that *the amplification factor $f_a$ is quite sensitively dependent on the age of galaxies as well as cosmological parameters.* The age of galaxies depends on the collapse time $z_{vir}$, and if the power spectrum of primeval density fluctuations, $|\delta_k|^2$ with a wavenumber $k$, is given by a power law, $|\delta_k|^2 \sim k^n$, then $z_{vir}$ is given as $(1 + z_{vir}) \propto M^{-(n+3)/6} \nu/b$, where $\nu$ is the amplitude of density perturbation in the units of rms density fluctuation, and $b$ denotes the biasing parameter for galaxy distribution. Thus, $z_{vir}$ is dependent on $M$, $\nu$, and $b$ as well as the shape of fluctuation spectrum. Although the determination of accurate values for $z_{vir}$ (i.e. for $z_{turn}$ and $z_f$), $H_0$, $\Omega$, and other quantities still eludes us, the magnetic field strength of protogalaxies at high redshifts can grow close to $\mu$G levels for some reasonable values for these parameters. The main reason for such strong magnetic fields at high redshifts is due to the combination of both the effect of dissipative galaxy collapse into a disk and the following exponential growth by non-axisymmetric gas flow.

We note that the standard dynamo models do *not* provide the dynamical time of a system as a time scale of field growth, but rather slow, of the order of 1Gyr, although the exact estimate depends on the unknown quantity, the so called $\alpha$-effect (see e.g. Zeldovich et al. 1983; Chiba & Lesch 1994).

The subsequent evolution of magnetic fields in galaxies is nonlinear, i.e. the magnetic fields are dynamically dominant in the gaseous turbulent motions or star-formation processes for angular-momentum braking. There is a limit for the field strength to satisfy the equilibrium condition $B^2/8\pi \sim \rho v^2/2$ where $\rho$ is the interstellar density ($\sim 1$cm$^{-3}$) and $v$ the random velocity ($\sim 8$kms$^{-1}$), giving $B \sim$ a few $\mu$G. Stronger fields in a disk should be inhibited by dynamical limit of local gas motion and flux expulsion due to Parker Instability. Although differential galactic rotation will continuously wrap up the field lines, the small but finite plasma drift with respect to neutral gas (i.e. ambipolar diffusion) will avoid the winding dilemma (Kulsrud 1986).

## 5. Concluding remarks

We have presented the evolution of magnetic fields in the context of galaxy formation. Three plasma processes have been considered, Compton drag, Battery effect, and sheared flow, as a creation of magnetic fields in a protogalactic over-dense region. Although each process gives only a very small field ranging $10^{-19} \sim 10^{-23}$G, the value adopted in the literature as a seed field, the following dissipative collapse into a disk strongly enhances the field strength by a factor $10^4$, or more, and leads to $10^{-15} \sim 10^{-19}$G depending on the redshift of disk formation. The strong non-axisymmetric disturbance due to bars or spirals in a protodisk will stretch and compress the magnetic-field lines with a dynamical time scale $\lesssim 10^8$yr. When we suppose $\Omega = 0.1$ and $z_f = 4$ as an epoch of galaxy formation, The amplification factor is $10^{10}$, leaving the magnetic-field strength $10^{-5} \sim 10^{-9}$G at $z \sim 2$, and more than $\mu$G at $z \sim 0.395$, possibility to explain the reported value of $\mu$G in damped Ly$\alpha$ clouds at high redshifts (Kronberg & Perry 1982; Kronberg et al. 1992; Wolfe et al. 1992). Therefore, the value of magnetic-field strength in galaxies can be explained by consistent consideration of plasma process in the early Universe, galaxy formation, and galactic dynamics.

The chain of distinctive processes for magnetic fields in galaxies as we have described is the simplification and idealization in order to elucidate the actual magnetic-field evolution step by step. This fact has been mentioned at the end of Sec.3, that each process can happen simultaneously. It is however our aim to clarify what is the most essential physics for magnetic fields in each stage of galaxy formation and evolution.

The battery can happen in "partially formed" galaxies which are in collapse phase. We have estimated the amplification factor by this process and found only three. It suggests that this effect can be negligible compared to the effect of collapse itself giving the much larger amplification factor of about $10^4$. The battery can also happen after a galaxy settled into centrifugal equilibrium. Though it is not clear to what extend and where (in a gaseous halo?) this process can work, it is already enough to invoke other field generation mechanisms such as dynamos. This is why the battery may be neglected after the collapse stage.

The battery as well as the shear type mechanism implies the existence of velocity shear in a pregalactic sphere. The inner region of a sphere, where the density is large and thus soon collapses, has obtained less torque from the surroundings than the outer region where the collapse time is large thus more time to get torque. It is thus natural to expect the appearance of velocity shear which would be of the order of angular velocity divided by the size. The assumption of an always rigidly rotating sphere is a very strong and artificial constraint on tidal torque theory.

We note that in the chain of every process before disk formation, the existence of ionized plasma is the assumption behind our analysis. In the stage of galaxy collapse, the baryon component is re-ionized due to virialization and thus collisional ionization. Also, before collapse, a pre-galactic gas is photoionized by UV flux from QSOs and other active galaxies which have formed before the epoch of normal galaxy formation (Efstathiou 1992; Chiba & Nath 1994). Thus it is not far from reality that the gas has existed as plasma.

Concerning the relative strength of field components in our model we note, that it is possible to know the pitch angle of field lines by the order of magnitude argument: In the case of the conventional galactic dynamos, the quantity $B_r/B_\phi$ can be written as

$$\frac{B_r}{B_\phi} \sim \left(\frac{R_\alpha}{R_\omega}\right)^{1/2}, \qquad (13)$$

where $R_\alpha = \alpha_0 h_0/\eta$ and $R_\omega = \omega_0 h_0^2/\eta$, $\eta$ denotes turbulent diffusion, and others hold usual meaning. In our model, this ratio is roughly written as

$$\frac{B_r}{B_\phi} \sim \frac{-R_{\phi\phi}\frac{B_r}{B_\phi} + R_{r\phi}}{R_\omega \frac{B_r}{B_\phi}}, \qquad (14)$$

where $R_{r\phi} = \frac{1}{r}\frac{\partial v_r}{\partial \phi}$ and $R_{\phi\phi} = \frac{1}{r}\frac{\partial v_\phi}{\partial \phi}$, both normalized by $\eta/h_0^2$. Though it is a rather complicated expression, we obtain the following two extreme cases: when $R_{r\phi}$ is dominant, $B_r/B_\phi \sim (R_{r\phi}/R_\omega)^{1/2}$; when $R_{\phi\phi}$ is dominant, $B_r/B_\phi \sim R_{\phi\phi}/R_\omega$. We note that this ratio directly depends on the relative strengths of each velocity shear driven by bars or spiral arms: for example, in case of M83, $R_{r\phi}$ is very large in bar regions so that the ratio is large, implying $B_r$ is dominant in such regions. This field configuration is such that the field lines are well aligned with a bar. In other words, our prediction concerning field pitch angles suggests the alignment of field lines with optical galactic structures.

Concerning the geometry of magnetic fields, we understand that it is also controversial even in galactic dynamo theory as to which field structure, ASS or BSS, a galaxy decides to hold as a dominant field geometry. Also, we understand that radio observers are also puzzled to interpret their observational data as to ASS or BSS. Thus we think that it is still premature to say about the field geometry. What we can learn from our model is that when non-axisymmetric disturbance in galaxy structure is dominant, it is the non-axisymmetric magnetic field that holds the largest growth rate.

Like conventional galactic dynamos, our mechanism for magnetic field generation considers the destructive process such as turbulent diffusion. When the combination of velocity shears attained in highly non-axisymmetric galaxies overcomes the destructive effects of turbulent diffusion, the strength of magnetic fields is increasing exponentially with time. This defines the threshold value for the combination of velocity shears, in our notation $\beta$, to give rise to the positive growth rate of magnetic fields. Thus in the sense that the field generation requires its overwhelming effect against turbulent diffusion, the mechanism we have quoted is similar to the conventional approach. However, as Hanasz and Lesch (1993) have shown, the Parker-instability may play an essential role for the magnetic diffusivity. With the assumption, that the magnetic field drives its own saturation via buoyant flux tubes ascending from the disk if the magnetic pressure is comparable to turbulent pressure in the disk, they obtained an $\alpha \simeq 1$ kms$^{-1}$, sufficient for dynamo action and a diffusivity $\eta_t \simeq 2 \cdot 10^{24}$ cm$^2$s$^{-1}$, which is three orders of magnitude smaller than the generally adopted value. Thus, the diffusion time is also much longer than in conventional dynamos. The picture of a magnetic diffusivity produced by magnetic instabilities instead of external driven diffusivity (via SN-explosion, stellar winds,etc...), has profound implications for the field amplification in protogalaxies. Since the field strengths in these objects is initially far away from the critical value for the Parker-instability, the fields are amplified until the magnetic pressure is comparable to the turbulent pressure in the protodisks. Then buoyant flux tubes start to transport magnetic flux into the halo and saturate the field amplification. The flux tubes can provide the $\alpha$-effect and an axisymmetric dynamo may start to act.

*Acknowledgements*. It is a pleasure to thank the referee Prof. A. Ruzmaikin for constructive criticisms and important suggestions. MC thanks the Alexander von Humboldt Foundation and the Max-Planck-Institut für Radioastronomie for support.

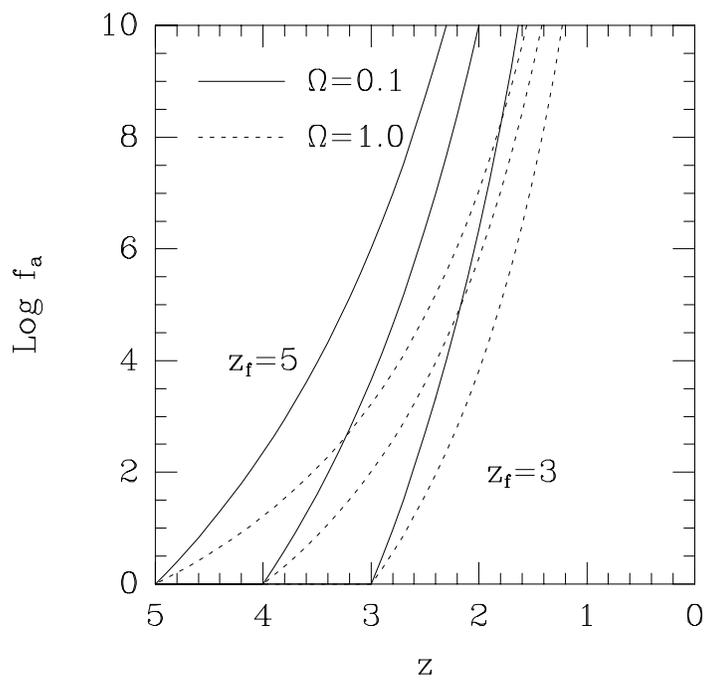